\documentclass[%
reprint,
superscriptaddress,
twocolumn,
amsmath,
amssymb,
aps,
prb
]{revtex4-1}

\usepackage{graphicx}  
\usepackage{dcolumn}  
\usepackage{bm}            
\usepackage{bbold}
\usepackage{amsmath}
\usepackage{xfrac}
\usepackage{xcolor}
\usepackage[colorlinks=true, citecolor=red]{hyperref}   
\usepackage{hhline}
\usepackage{diagbox}
\usepackage{siunitx}
\usepackage{enumitem}
\usepackage[normalem]{ulem}

\newcommand{\kelvin}{\textrm{~K}}
\newcommand{\kOhm}{~\textrm{\SI{}{\kohm}}}
\newcommand{\Ohm}{~\textrm{\SI{}{\ohm}}}
\newcommand{\simm}{$\sim$}
\usepackage{siunitx}

\usepackage{tabularx}
\usepackage{array} 

\newcolumntype{Y}{>{\centering\arraybackslash}X}

\begin{document}
	
	\title{Integer and fractional quantum anomalous Hall effects in pentalayer graphene}

	\author{Ming Xie}
	\affiliation{Condensed Matter Theory Center and Joint Quantum Institute, Department of Physics, 
		University of Maryland, College Park, Maryland 20742, USA}
	
	\author{Sankar Das Sarma}
	  	    \affiliation{Condensed Matter Theory Center and Joint Quantum Institute, Department of Physics, 
			                University of Maryland, College Park, Maryland 20742, USA}
      
	\date{\today}
	
	\begin{abstract}
		
	We critically analyze the recently reported observation of
	 integer (IQAHE) and fractional (FQAHE) quantum anomalous Hall  effects 
	at zero applied magnetic field in pentalayer graphene.  
	Our quantitative activation and variable range hopping transport analysis of the experimental data reveals that the observed IQAHE and FQAHE at different fillings all have similar excitation gaps of the order of $5$\kelvin$-10$\kelvin.  
	In addition, we also find that the observed FQAHE manifests a large hidden background contact series resistance $>$10\kOhm~of unknown origin whereas this contact resistance is much smaller \simm500\Ohm\ in the observed IQAHE.  
	Both of these findings are surprising as well as inconsistent with the well-established phenomenology of the corresponding high-field integer and fractional quantum Hall effects in 2D semiconductor systems.

	\end{abstract}
	
	\maketitle

The recent transport observation of zero-field integer (IQAHE) and fractional (FQAHE) quantum anomalous Hall effects in pentalayer graphene, \cite{PentaGraphene2023} just after the report of the same discovery in twisted MoTe$_2$, \cite{Park2023,Fan2023} is an important breakthrough. 
Graphene is relatively clean, allowing the possibility of seeing novel quantum Hall effect (QHE) physics, perhaps not necessarily accessible in dirtier TMD materials.  
It should be mentioned as an aside that the original high-field IQHE was first observed  \cite{Klitzing1980} in 2D Si-SiO$_2$ MOSFETs with low mobilities (\simm$10^3$ cm$^2$/Vs) 
whereas the first observation \cite{Tsui1982} of FQHE was in much cleaner 2D GaAs-AlGaAs heterostructures with much higher mobilities (\simm$10^5$ cm$^2$/Vs).  
To the best of our knowledge, FQHE has yet to be decisively reported in any transport experiments on Si-SiO$_2$ MOSFETs because of their high disorder.
It is therefore remarkable that the zero field version of quantum Hall effects manifested simultaneously as IQAHE and FQAHE in both TMD and graphene.  
Indeed, Ref.~\onlinecite{PentaGraphene2023} reports the observation of ‘plateaus’ of quantized Hall resistance $R_{xy} =h/\nu e^2$ at the integer band filling $\nu=1$ as well as several fractional band fillings $\nu = 2/3, 3/5, 2/5, 4/9, 3/7$ and $4/7$ accompanied by clear dips in the longitudinal resistance.
The current work follows our earlier work on TMD \cite{DasSarma2024}, and focuses on a detailed analysis of the experimental IQAHE/FQAHE results presented in Ref.~\onlinecite{PentaGraphene2023}.

The qualitative differences between regular high-field QHEs and the recently observed IQAHE/FQAHE are as follows: 
(1) high magnetic field versus zero magnetic field; 
(2) magnetic field induced continuum Landau levels (LLs) versus lattice induced flat-bands with nontrivial topology 
(i.e. Chern insulators with a nontrivial Chern number); 
(3) explicit breaking of the time reversal symmetry by an external magnetic field versus the spontaneous breaking of the time reversal symmetry in a Chern band.  
The interplay of continuum LLs and a lattice in the context of IQHE/FQHE has a long history going back to the early seminal papers.\cite{Thouless1982, Haldane1988, Kol1993}
Later, ideas were proposed to create FQHE in atomic optical lattices.\cite{Lukin2005} 
Occasionally, specific models for creating flat bands, which are necessary (but not sufficient) for IQAHE, 
in 2D lattices were discussed, but without any embedded topology or time reversal symmetry breaking.\cite{Wu2007}
Finally, in 2011, several theoretical groups proposed 2D lattice models where the bands are both relatively flat and have intrinsic topology (i.e. finite Chern number), 
leading to the specific idea of quantum Hall effects without any applied field.  
\cite{Tang2011, Sun2011, Neupert2011, Regnault2011, Sheng2011} 
It was obvious right from the beginning (and  also verified by explicit numerical simulations) that such Chern bands, when fractionally occupied, would manifest FQAHE.
The possibility of
Chern insulators with Chern numbers larger than one 
and the role of disorder in IQAHE/FQAHE in such Chern insulators were also considered in this early theoretical literature.
\cite{Shuo2012a, Shuo2012b}
More recently, a specific prediction for the occurrence of fractional Chern insulators in twisted 2D moir\'e systems was made. \cite{Li2021}
However, the subject remained strictly theoretical 
without any experimental realization for 12 years until 2023 
when three groups realized Chern insulators with observations of IQAHE/FQAHE in TMD and pentalayer graphene moir\'e systems in transport experiments.
\cite{Park2023, Fan2023, PentaGraphene2023}  

Given that IQAHE/FQAHE phenomena in flatband Chern insulators were already predicted as a matter of principle a long time ago, their observations bring up three important theoretical questions:  
(1) is the observed zero-field IQAHE/FQAHE in 2D lattice systems adiabatically connected to the corresponding well-established high-field IQHE/FQHE in continuum LL systems? 
(2) can the observed experimental IQAHE/FQAHE phenomenology theoretically quantitatively explicable using the actual TMD and pentalayer graphene lattices?  
(3) are there new fractions where FQAHE manifest with no corresponding FQHE analogs?
Our current Letter illuminates on these questions by critically analyzing the reported pentalayer graphene data in depth 
and showing that the reported IQAHE/FQAHE in graphene has serious quantitative disagreement with the corresponding high-field IQHE/FQHE phenomenology, 
indicating that perhaps the observed IQAHE/FQAHE may \textit{not} necessarily be adiabatically connected to the high field quantum Hall physics.  
This is, at this early stage of the subject, at best a tentative conclusion since experimental results are likely to improve rapidly as sample quality improves and experimental problems are resolved,
but the subject is of sufficient importance for the community to think about this issue.  

We emphasize in this context that no theoretical work\cite{Senthil2023,JHU2023,Parker2023, BernevigII2023, BernevigIII2023, Liu2023, Parker2024}
to date provides even a semi-quantitative explanation for the experimental observations of the pentalayer graphene IQAHE/FQAHE.
For example, theories either find FQAHE both
\cite{Senthil2023,JHU2023,Parker2023,Liu2023} or neither \cite{BernevigPrivate} 
for $\nu=1/3$ and $2/3$ in graphene whereas Ref.~\onlinecite{PentaGraphene2023} reports the $\nu=2/3$, 
but not the $\nu=1/3$ FQAHE.  
\begin{figure}[t!]
	\includegraphics[width=0.4\textwidth]{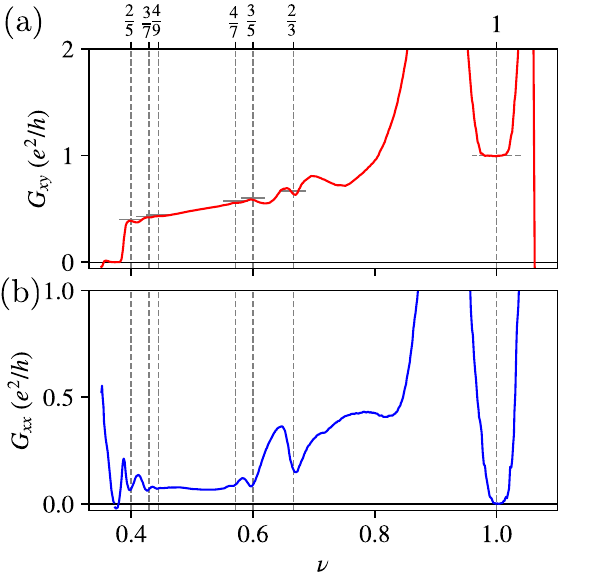}
	\caption{\label{sigma} 
		(a) Hall and (b) longitudinal conductances of the pentalayer graphene/hBN as a function of filling factor $\nu$.
		$G_{xy}$ and $G_{xx}$ are obtained by inverting the resistances data from the source data of Fig.~3(c) in Ref.~\onlinecite{PentaGraphene2023}. 
		The vertical dashed lines represent the fillings manifesting the IQAHE/FQAHE,
		and the horizontal bars mark the expected quantized Hall conductance values.
		Note that the measured quantities are the resistances, and our fitting is done on $R_{xx}(T)$.
	}
\end{figure}
At this stage we do not know whether or not the reported existence (nonexistence) of $\nu =2/3 (1/3)$ 
FQAHE is an unimportant detail or not.
On the other hand, Ref.~\onlinecite{PentaGraphene2023} reports FQAHE precisely at the primary Jain fractional fillings  
arising from the composite fermion theory (except for 1/3 filling) as developed for the LL FQHE.

In the rest of this Letter we present our detailed analysis of 
the temperature dependent longitudinal resistance $R_{xx}$ 
using the experimental data in Ref.~\onlinecite{PentaGraphene2023} 
to obtain the IQAHE/FQAHE excitation gaps at various fillings.  
We also find that, similar to the situation in TMD IQAHE/FQAHE \cite{DasSarma2024},
pentalayer graphene data also imply a hidden (and often large) series resistance $R_0$ in the longitudinal resistance, 
whose origin remains a mystery (since 4-probe measurements should not usually manifest any contact resistance).  
We provide the excitation gaps and the ‘contact’ resistance for all the fillings 
where Ref.~\onlinecite{PentaGraphene2023} reports the observation of an apparent quantization of $R_{xy}$.

\begin{figure}[b!]
	\includegraphics[width=0.45\textwidth]{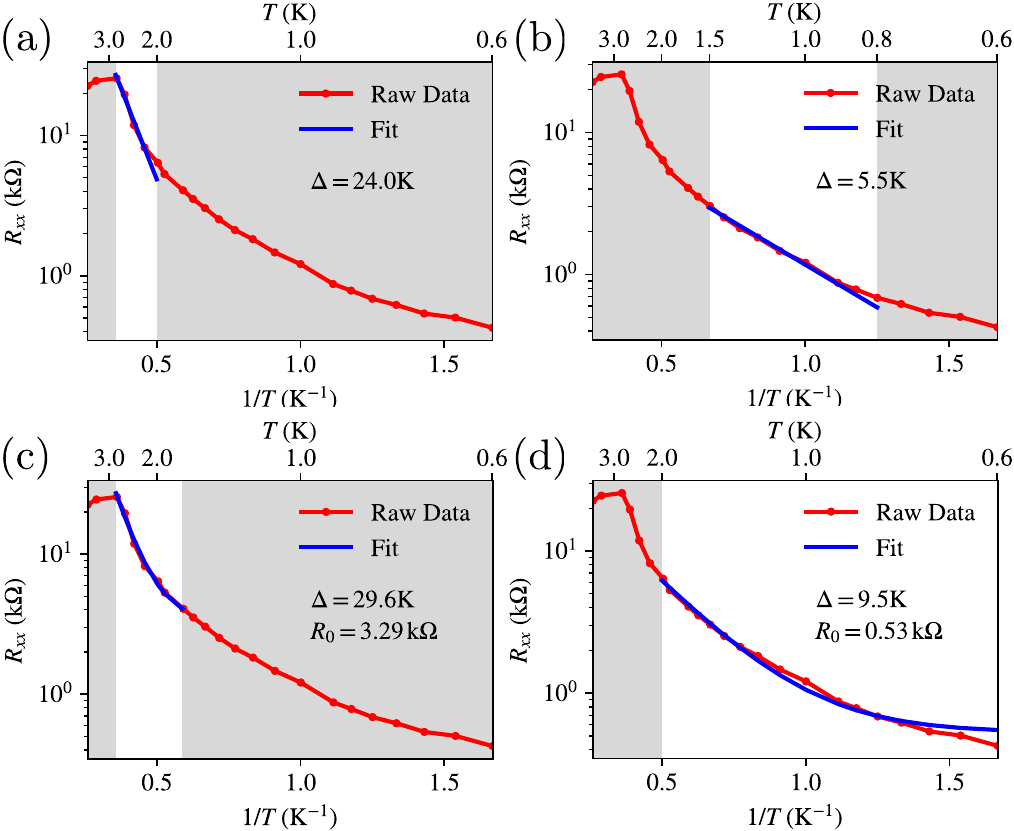}
	\caption{\label{int_activation} Thermal activation fitting of the longitudinal resistance $R_{xx}$ at filling factor $\nu=1$. 
		(a,b) are fittings without contact resistance and 
		(c,d) are fittings with a contact resistance $R_0$ as a fitting parameter. 
		The red lines are experiment data and blue lines are the fitting curves. 
		Data in the shaded region is excluded from fitting.
	}
\end{figure}

We start from an overall examination of the degree of quantization for both the integer and fractional filling states.
Figure~\ref{sigma} plots the the Hall ($G_{xy}$)  and longitudinal ($G_{xx}$) conductances as a function of filling factor.
The conductances are calculated from the resistance data (shown in Fig.~3(c) of Ref.~\onlinecite{PentaGraphene2023}) measured in a constant current setup by $G_{xx/xy}=R_{xx/xy}/(R_{xx}^2+R_{xy}^2)$.
While both the integer and fractional plateaus are accompanied by clear dips in $G_{xx}$,
the residual resistance (and thereby the conductance) is significant ($\sim10$\kOhm) in the fractional cases, 
several orders of magnitude higher compared to the integer case, which can reach as low as $\sim10\ \Omega$\cite{PentaGraphene2023}.
In fact, the surprise here is why and how the FQAHE shows up at 6 different fractional fillings in spite of a huge background resistance---this certainly does not happen in regular FQHE.
The existence of a large ($\sim10$\kOhm) residual resistance concomitant with the manifestation of clear FQAHE 
sharply distinguishes the lattice FQAHE from LL FQHE, 
and this calls for a deeper understanding of the relative roles of edge versus bulk transport in FQAHE to figure out what this mysterious $R_0$ is.

To have a quantitative understanding, 
we focus on the temperature dependence of the longitudinal resistance $R_{xx}$.
In ideal clean systems, because of the ballistic transport along the chiral edge channels,
the longitudinal resistance is expected to vanish in the limit of zero temperature.
As temperature rises, thermally activated carriers in the bulk contribute
to finite resistance,
which follows the thermal activation behavior with $R_{xx} \propto e^{-\Delta/2k_BT} $,
where $\Delta$ is the bulk charge excitation gap.
Additional mechanisms contributing to $R_{xx}$ are 
Mott \cite{Mott1969}  and Efros-Shklovskii (ES)  \cite{ES1975} variable range hopping (VRH) facilitated by phonons, 
and the relative importance of the three transport mechanisms in IQAHE/FQAHE is unknown.

Here we fit the experimental temperature dependence of $R_{xx}$ to all three transport mechanisms 
in an unbiased manner.
In addition, we include an extra term, the contact resistance $R_0$, which was found to be essential in the TMD
IQAHE/FQAHE \cite{DasSarma2024}.
We remark that the exact origin of $R_0$ is unclear at the moment; 
the fact that $R_0$ exists is apparent in the experimental data since the measured $R_{xx}$, instead of becoming vanishingly small at the FQAHE plateau, saturates to a large value.
Specifically, we adopt the fitting formula $R(T) = R_0 + Ae^{-\Delta/2k_BT} $ for the thermal activation mechanism,
and $R(T) = R_0 + Ae^{-(T_0/T)^{1/n}}$ for Mott ($n=3$) and ES ($n=2$) VRH mechanisms,
where $A$ and $T_0$ are model dependent constants treated as free fitting parameters.
Below we present our fitting results for filling factors $\nu=1, 2/3, 3/5, 2/5$, and $4/9$.
We provide our fitted results for $\nu=3/7$ and $4/7$ also (see Table~\ref{summary}), 
but do not show the actual fittings for brevity.

\begin{figure}[t!]
	\includegraphics[width=0.41\textwidth]{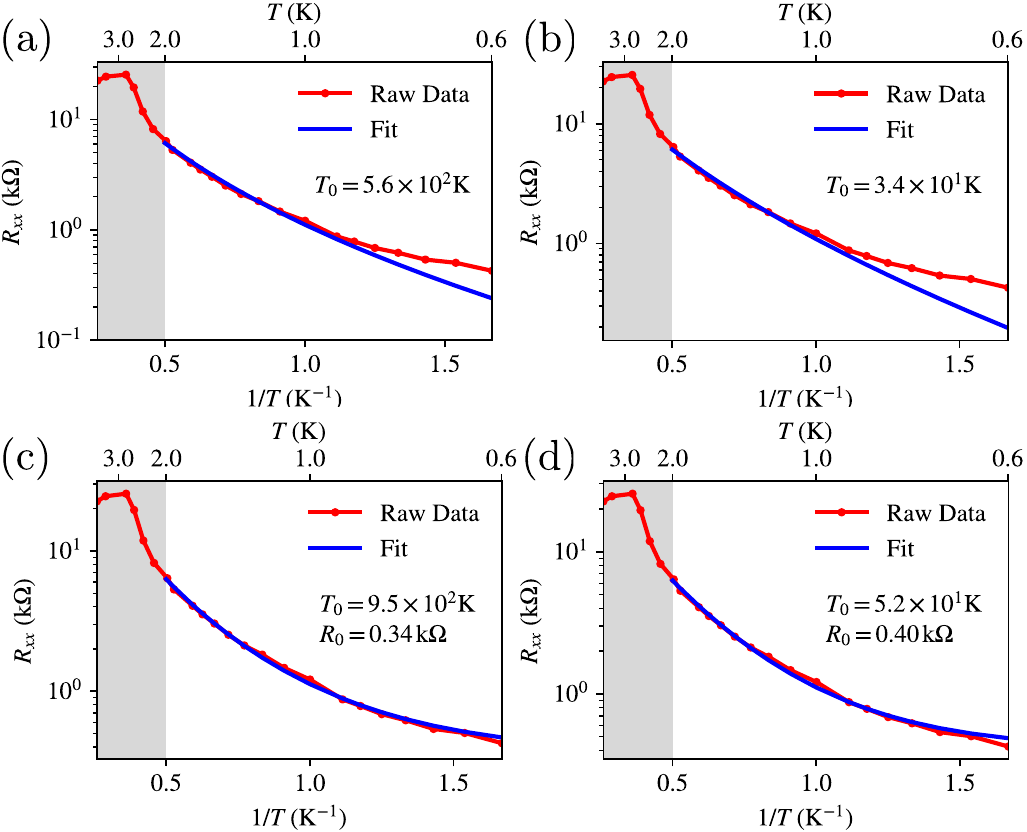}
	\caption{\label{int_VRH} 
		Variable range hopping fittings of the longitudinal resistance $R_{xx}$ at filling factor $\nu=1$.
		The left panels are Mott VRH fittings without (a) and with (c) a contact resistance $R_0$.
		The right panels are ES VRH fittings without (b) and with (d) a contact resistance $R_0$.
		The figure conventions are the same as in Fig.~\ref{int_activation}.
	}
\end{figure}

At the integer filling $\nu=1$, $R_{xx}$ increases with increasing temperature (or decreasing $T^{-1}$)
and saturates at a temperature of $T_s\approx 3\ $K as shown in Fig.~\ref{int_activation}.
Below $T_s$, thermal activation fitting without $R_0$ fails to match the overall temperature dependence, as
shown in Fig.~\ref{int_activation} (a-b), with only piecewise matchings 
yielding drastically different gap values depending crucially on the fitting range.
With a finite $R_0$, the fitting appears to have a significantly better agreement with the measured data
as demonstrated in Fig.~\ref{int_activation}(d).
The temperature range close to $T_s$ manifests a different behavior (shown in Fig.~\ref{int_activation}(c)) likely due to the saturation effect.
The residual resistance is generally small but can vary from 
$\sim100\ \Omega$ to $\sim10\ \Omega$\cite{PentaGraphene2023}
depending on detailed parameters such as the vertical displacement field.
The extracted value for the gap is $\Delta\approx9.5$ K.
We emphasize that our extracted (rather small) IQAHE gap ($\sim9.5$ K) is consistent with the experimental finding 
that the IQAHE disappears for $T>2\ K$ which argues against a large gap.  
Similarly, the measured  $R_{xx}$ at the lowest temperatures is $\sim10-100$\Ohm\ at the IQAHE plateau in Ref.~\onlinecite{PentaGraphene2023} consistent with our fitted $R_0$.

\begin{figure}[t!]
	\includegraphics[width=0.45\textwidth]{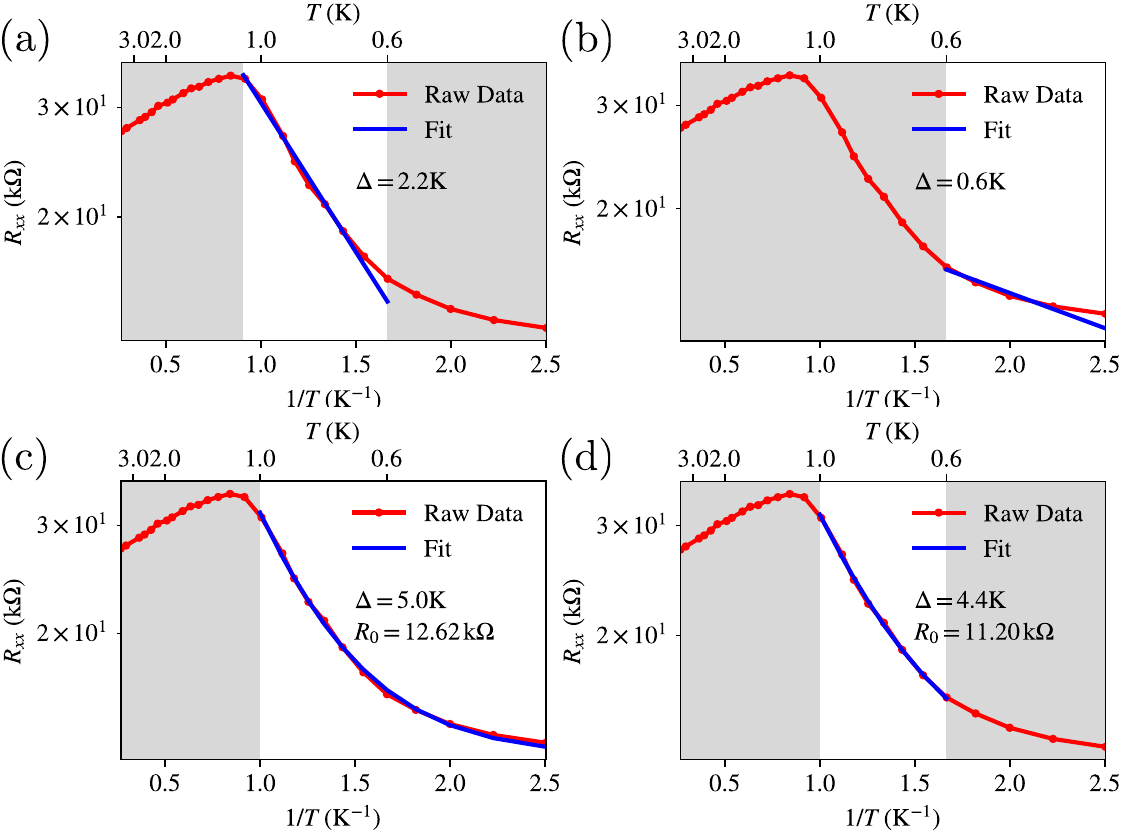}
	\caption{\label{23activation} Thermal activation fitting of the longitudinal resistance $R_{xx}$ at filling factor $\nu=2/3$.
		(a,b) are fittings without contact resistance and 
		(c,d) are fittings with a contact resistance $R_0$. 
		The figure conventions are the same as in Fig.~\ref{int_activation}.
	}
\end{figure}

We further apply the Mott and ES VRH mechanisms to fit the temperature dependence of $R_{xx}$ at $\nu=1$.
Figure~\ref{int_VRH} presents the optimal fitting results (a,b) without and (c,d) with a contact resistance,
which clearly demonstrates the necessity to include $R_0$.
The fact that the VRH and thermal activation mechanisms can both fit the experimental data
suggests the mechanism of finite temperature $R_{xx}$ is uncertain.

For FQAHE, the longitudinal resistances $R_{xx}$ have similar trends
but differ quantitatively compared to the integer case.
We begin with the filling factor $\nu=2/3$, which is the most robust fractional state.
As shown in Fig.~\ref{23activation}, $R_{xx}$ first increases and then decreases as the temperature $T$ increases.
Similar to the integer case, inclusion of $R_0$ is necessary to fit the temperature dependence.
We find only minor variations in the thermal activation fitting parameters when adjusting the temperature window 
(as shown in Fig.~\ref{23activation}(c,d))\cite{footnote1}.
The optimal gap value from our fitting is $\Delta=5.0$ K, approximately half of that for the integer case.
The contact resistance $R_0=12.62$\kOhm, on the other hand, is 2 orders of magnitude larger than the integer case.
Our extracted fractional gap $\sim5$ K and $R_0\sim10$\kOhm\ are consistent with the experimental FQAHE 
disappearing below 1 K and the experimental $R_{xx}$ manifesting a resistance of $\sim10$\kOhm\ 
around the fractional plateau at the lowest temperatures, respectively.

\begin{figure}[t!]
	\includegraphics[width=0.45\textwidth]{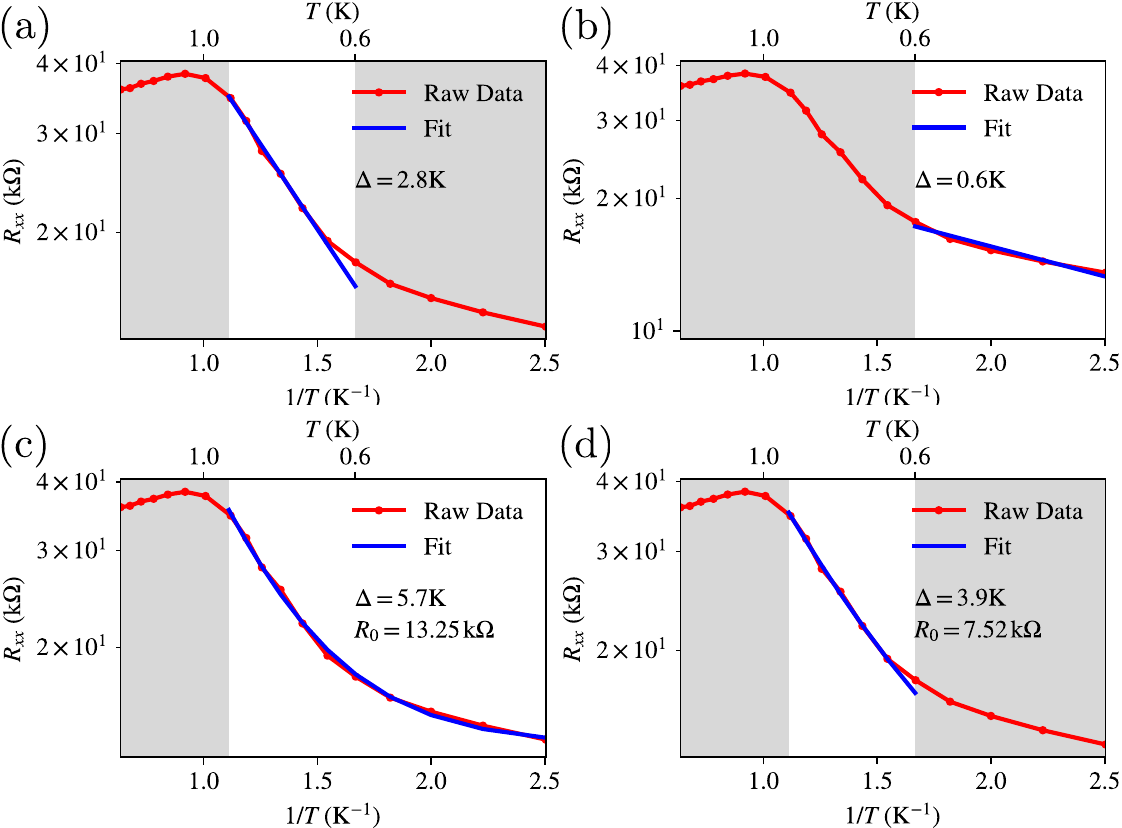}
	\caption{\label{35activation} Thermal activation fitting of the longitudinal resistance $R_{xx}$ at filling factor $\nu=3/5$.
		(a,b) are fittings without contact resistance and 
		(c,d) are fittings with a contact resistance $R_0$. 
		The figure conventions are the same as in Fig.~\ref{int_activation}.
	}
\end{figure}
\begin{figure}[t!]
	\includegraphics[width=0.45\textwidth]{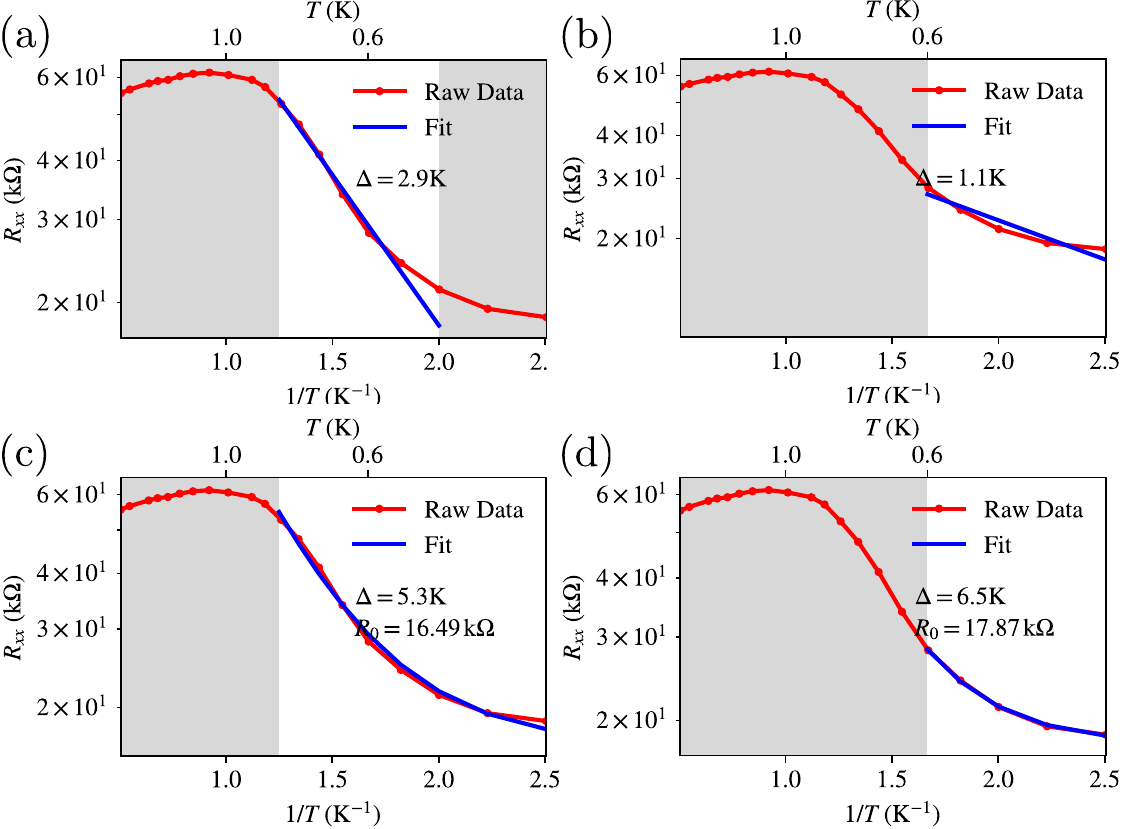}
	\caption{\label{25activation} 
		Thermal activation fitting of the longitudinal resistance $R_{xx}$ at filling factor $\nu=2/5$. 
		(a,b) are fittings without contact resistance and 
		(c,d) are fittings with a contact resistance $R_0$ as a fitting parameter. 
		The figure conventions are the same as in Fig.~\ref{int_activation}.
	}
\end{figure}
\begin{figure}[t!]
	\includegraphics[width=0.45\textwidth]{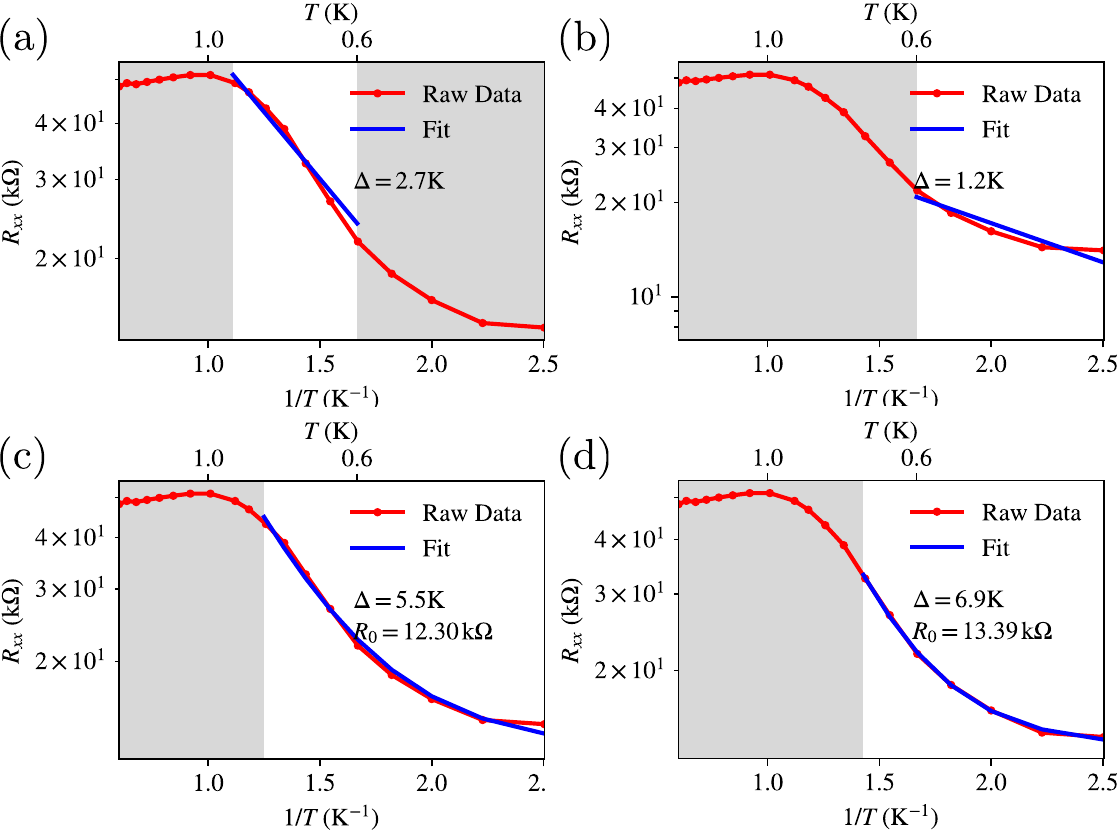}
	\caption{\label{49activation} 
		Thermal activation fitting of the longitudinal resistance $R_{xx}$ at filling factor $\nu=4/9$. 
		(a,b) are fittings without contact resistance and 
		(c,d) are fittings with a contact resistance $R_0$ as a fitting parameter. 
		The figure conventions are the same as in Fig.~\ref{int_activation}.
	}
\end{figure}

We perform similar thermal activation fitting analysis for several other fractional filling factors.
Figure~\ref{35activation}, \ref{25activation} and \ref{49activation} show the fitting results for
$\nu=3/5, 2/5$ and $4/9$ respectively ($\nu=3/7$ and $4/7$ are not shown).
The longitudinal resistances in these cases follow similar trend with a saturation (or turnover) 
temperature $T_s\approx 1$ K.
Again, for $T<T_s$, thermal activation fitting with finite $R_0$ works well in all cases
yielding similar $R_0$ values $\sim10$ \kOhm.
Table~\ref{summary} summarizes the fitting parameters from all the thermal activation fittings.
We remark that our fitting method aligns with standard practices in the extensive FQHE literature. 
Although using all three transport mechanisms yields somewhat better fits, 
it involves too many parameters, making the fitting less meaningful.

\begin{table}[b!] 
	\caption{Summary of the fitting parameters for thermal activation fitting of the longitudinal resistance $R_{xx}$.
		The parameters are extracted from the optimal fitting result at each filling factor. 
		$\Delta$ and $R_0$ are in units of Kelvin (K) and \textrm{\SI{}{\kohm}} respectively.}
	\label{summary}
	\begin{tabularx}{.48\textwidth}{YYYYYYYY}
		\hline\hline
		$\nu$ & 1 & 2/3 & 3/5 & 2/5 &  4/9 & 3/7 & 4/7\\ 
		\hline
		$\Delta$                               & 9.5   & 5.0      & 5.7   & 5.3     & 5.5  & 4.9 & 5.0 \\ 
		$R_0$   & 0.53 & 12.62  & 13.25   & 16.49    & 12.30 & 12.12 & 10.29 \\ 
		\hline\hline
	\end{tabularx}
	
\end{table}

The most interesting and important aspect of our results, summarized in Table~\ref{summary}, 
is that the FQAHE gap is consistently around $5$ K for all the fractional fillings, 
and the IQAHE gap is within a factor of 2 of the FQAHE gap.  
However, the experimental quantization is far better for $\nu=1$ than for the fractional fillings, 
which is likely caused by the effective background contact resistance $R_0$ 
being orders of magnitude larger for the fractional cases ($>10 $\kOhm) than in the integer case ($\sim10-100$\Ohm).  
We mention that if we blindly convert $R_0\sim10$ \kOhm\ to a broadening 
using the actual carrier density and the very light carrier effective mass ($\sim0.05 m_e$) of pentalayer graphene, 
we get a huge broadening $\sim500$ K. 
The actual transport broadening is, however, likely to be much smaller $\sim10$ K 
as reflected from the onset of the Shubnikov–de Haas (SdH) oscillations at $0.4$ T applied field. 
This $10$ K broadening is consistent with the IQAHE value of $R_0\sim100$~\Ohm.

\begin{figure*}
	\includegraphics[width=0.99\textwidth]{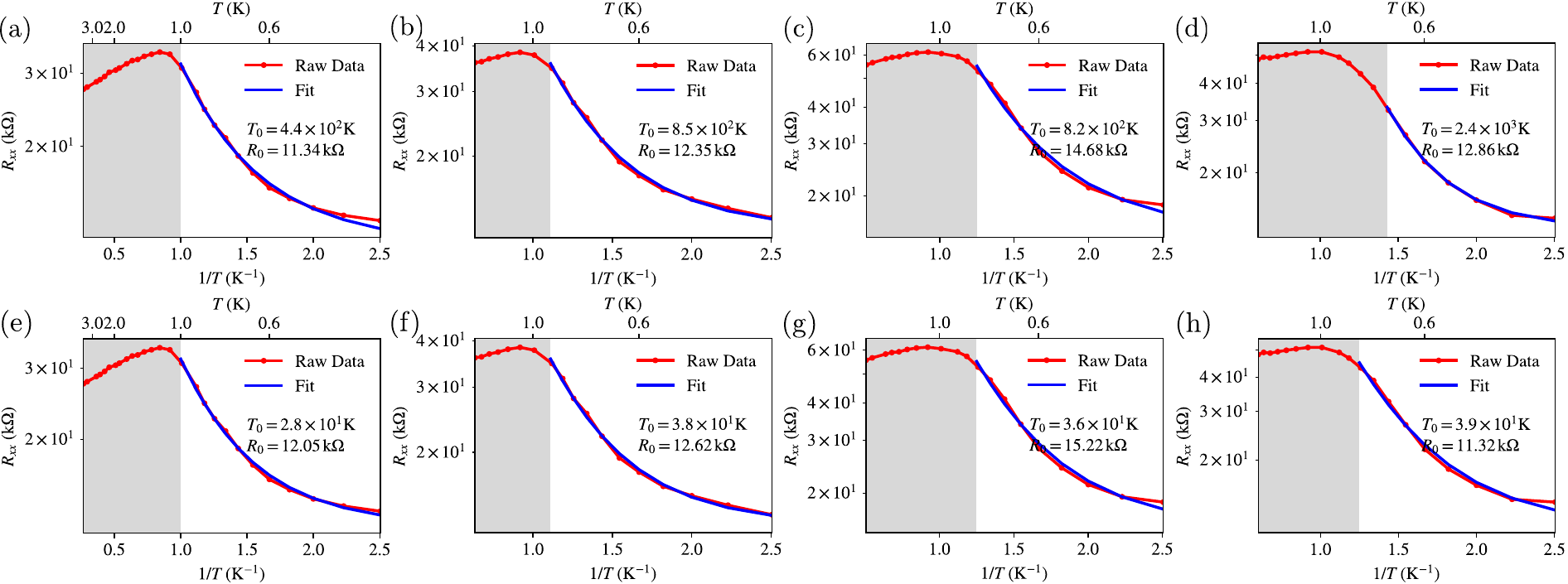}
	\caption{\label{fracVRH} (a-d) Mott and (e-h) ES VRH fittings of the longitudinal resistance $R_{xx}$ 
		at fractional filling factors with (a,e)  for $\nu=2/3$, (b,f) for  $\nu=3/5$,  (c,g) for  $\nu=2/5$,  and (d,h) for  $\nu=4/9$.
		The figure conventions are the same as in Fig.~\ref{int_activation}.
	}
\end{figure*}

The biggest puzzle is perhaps the almost constancy of the gap values across all observed FQAH states.  
Since the observed fractions are precisely (except for the absence of the 1/3 state) the primary Jain sequence
 $\nu=p /(2p+1)$ or $(p+1)/(2p +1)$ with $p=1,2,3....$, 
exactly as occurring in the continuum high-field 2D GaAs LL FQHE experiments,\cite{West1993, Shayegen2022, Shayegen2021}
it is reasonable to assume an adiabatic continuity between the continuum LL and the flatband lattice systems.  
By contrast, however, the composite fermion theory in the continuum system, 
which is well-verified both experimentally \cite{Shayegen2022, Shayegen2021}
and theoretically,\cite{Morf2002, Jain2002}
specifically predicts that the excitation gap should scale as $1/(2p+1)$
whereas the experimental pentalayer graphene FQAH gaps in Table I show no dependence on $p$ (or $\nu$).  
For example, the composite fermion theory \cite{HalperinBook,JainBook}
as  applied to the current graphene FQAHE would predict the gap at $2/3$ to be three times the gap at $4/9$
(and this is exactly what is observed experimentally in LL FQHE \cite{Shayegen2021})
whereas our FQAHE analysis in Table~\ref{summary} finds the gap at $4/9$ to be slightly larger.  
It is possible that the flatband lattice FQAHE represents a distinct type of topological order 
different from its continuum LL counterpart since the lattice version, in principle, 
has many additional topological invariants imposed by lattice translational symmetries. 
\cite{Barkeshli2012,Barkeshli2021}
It is too soon to decide whether some deep significance should be attached to the excitation gap 
being almost a constant for all the FQAH states (Table~\ref{summary}) 
or this is simply an artifact of the system being highly disordered,
particularly since the observed FQAHE sequence in graphene is 
precisely the same as the primary Jain sequence predicted for the continuuum FQHE. 
We do, however, believe that our Table~\ref{summary} represents an important empirical fact about FQAHE 
which future theories must be able to explain.  

We also perform VRH fittings for the fractional cases considered above.
The fitting results are shown in Fig.~\ref{fracVRH} with the Mott fittings in the top row
and ES fittings in the bottom row.
Both VRH models are able to fit the $R_{xx}$ data.
The extracted $R_0$ values are consistent with the thermal activation results in Table~\ref{summary}.
It is clear from these results that we cannot unambiguously pin down the nature of the transport mechanism 
and therefore should be cautious about adopting the gap value extracted from the thermal activation fitting,
accepting them as tentative findings until future results in better samples with much lower values of $R_0$ are reported.

\begin{acknowledgments}
	{\em Acknowledgment.}---\noindent	The authors thank Long Ju and Zhengguang Lu for providing the temperature dependent numerical data corresponding to the published results in Ref.~\onlinecite{PentaGraphene2023} and for helpful discussions.  This work is supported by Laboratory for Physical Sciences.
\end{acknowledgments}


\begin{thebibliography}{99}

\bibitem{PentaGraphene2023}
Z. Lu, T. Han, Y. Yao, A. P. Reddy, J. Yang, J. Seo, K. Watanabe, T. Taniguchi, L. Fu, and L. Ju,
Fractional quantum anomalous Hall effect in a graphene moir\'e superlattice,
Nature (London) \textbf{626}, 759 (2024).

\bibitem{Park2023}
H. Park, J. Cai, E. Anderson, Y. Zhang, J. Zhu, X. Liu, C. Wang, W. Holtzmann, C. Hu, Z. Liu, T. Taniguchi, K. Watanabe, J.-H. Chu, T. Cao, L. Fu, W. Yao, C.-Z. Chang, D. Cobden, D. Xiao, and X. Xu,
Observation of fractionally quantized anomalous Hall effect,
Nature  (London) \textbf{622}, 74 (2023).

\bibitem{Fan2023}
F. Xu, Z. Sun, T. Jia, C. Liu, C. Xu, C. Li, Y. Gu, K. Watanabe, T. Taniguchi, B. Tong, J. Jia, Z. Shi, S. Jiang, Y. Zhang, X. Liu, and T. Li,
Observation of Integer and Fractional Quantum Anomalous Hall Effects in Twisted Bilayer MoTe$_2$,
Phys. Rev. X \textbf{13}, 031037 (2023).

\bibitem{Klitzing1980}
K. v. Klitzing, G. Dorda, and M. Pepper,
New Method for High-Accuracy Determination of the Fine-Structure Constant Based on Quantized Hall Resistance,
Phys. Rev. Lett. \textbf{45}, 494 (1980).

\bibitem{Tsui1982}
D. C. Tsui, H. L. Stormer, and A. C. Gossard,
Two-Dimensional Magnetotransport in the Extreme Quantum Limit,
Phys. Rev. Lett. \textbf{48}, 1559 (1982).

\bibitem{DasSarma2024}
S. Das Sarma and M. Xie,
On the zero-field quantization of the anomalous quantum Hall effect in two-dimensional moir\'e layers,
Phys. Rev. B \textbf{109}, L121104 (2024).

\bibitem{Thouless1982}
D. J. Thouless, M. Kohmoto, M. P. Nightingale, and M. den Nijs,
Quantized Hall conductance in a two-dimensional periodic potential,
Phys. Rev. Lett. \textbf{49}, 405 (1982).

\bibitem{Haldane1988}
F. D. M. Haldane,
Model for a Quantum Hall Effect without Landau Levels: Condensed-Matter Realization of the ``Parity Anomaly",
Phys. Rev. Lett. \textbf{61}, 2015 (1988).

\bibitem{Kol1993}
A. Kol and N. Read,
Fractional quantum Hall effect in a periodic potential,
Phys. Rev. B \textbf{48}, 8890 (1993).

\bibitem{Lukin2005}
A. S. Sørensen, E. Demler, and M. D. Lukin,
Fractional Quantum Hall States of Atoms in Optical Lattices,
Phys. Rev. Lett. \textbf{94}, 086803 (2005).

\bibitem{Wu2007}
C. Wu, D. Bergman, L. Balents, and S. Das Sarma,
Flat Bands and Wigner Crystallization in the Honeycomb Optical Lattice,
Phys. Rev. Lett. \textbf{99}, 070401 (2007).


\bibitem{Tang2011}
E. Tang, J.-W. Mei, and X.-G. Wen,
High-Temperature Fractional Quantum Hall States,
Phys. Rev. Lett. \textbf{106}, 236802 (2011).

\bibitem{Sun2011}
K. Sun, Z. Gu, H. Katsura, and S. Das Sarm,
Nearly Flatbands with Nontrivial Topology,
Phys. Rev. Lett. 106, 236803 (2011).

\bibitem{Neupert2011}
T. Neupert, L. Santos, C. Chamon, and C. Mudry,
Fractional Quantum Hall States at Zero Magnetic Field,
Phys. Rev. Lett. \textbf{106}, 236804 (2011).


\bibitem{Regnault2011}
N. Regnault and B. A. Bernevig,
Fractional Chern Insulator,
Phys. Rev. X \textbf{1}, 021014 (2011).

\bibitem{Sheng2011}
D. N. Sheng, Z.-C. Gu, K. Sun, and L. Sheng,
Fractional quantum Hall effect in the absence of Landau levels,
Nat. Comm. \textbf{2}, 389 (2011).

\bibitem{Shuo2012a}
S. Yang, K. Sun, and S. Das Sarma,
Quantum phases of disordered flatband lattice fractional quantum Hall systems,
Phys. Rev. B \textbf{85}, 205124 (2012).

\bibitem{Shuo2012b}
S. Yang, Z.-C. Gu, K. Sun, and S. Das Sarma,
Topological flat band models with arbitrary Chern numbers,
Phys. Rev. B \textbf{86}, 241112(R) (2012).


\bibitem{Li2021}
H. Li, U. Kumar, K. Sun, and S.-Z. Lin,
Spontaneous fractional Chern insulators in transition metal dichalcogenide moir\'e superlattices,
Phys. Rev. Research \textbf{3}, L032070 (2021).

 
 \bibitem{Senthil2023}
 Z. Dong, A. S. Patri, T. Senthil,
 Theory of fractional quantum anomalous Hall phases in pentalayer rhombohedral graphene moiré structures, arXiv:2311.03445.
 
 \bibitem{JHU2023}
 B. Zhou, H. Yang, Y.-H. Zhang,
 Fractional quantum anomalous Hall effects in rhombohedral multilayer graphene in the moiréless limit and in Coulomb imprinted superlattice, arXiv:2311.04217.
 
 \bibitem{Parker2023}
 J. Dong, T. Wang, T. Wang, T. Soejima, M. P. Zaletel, A. Vishwanath, D. E. Parker,
 Anomalous Hall Crystals in Rhombohedral Multilayer Graphene I: Interaction-Driven Chern Bands and Fractional Quantum Hall States at Zero Magnetic Field, arXiv:2311.05568.
 
 \bibitem{BernevigIII2023}
 Y. H. Kwan, J. Yu, J. Herzog-Arbeitman, D. K. Efetov, N. Regnault, B. Andrei Bernevig,
 Moiré Fractional Chern Insulators III: Hartree-Fock Phase Diagram, Magic Angle Regime for Chern Insulator States, the Role of the Moiré Potential and Goldstone Gaps in Rhombohedral Graphene Superlattices, arXiv:2312.11617.
 
 \bibitem{BernevigII2023}
 J. Herzog-Arbeitman, Y. Wang, J. Liu, P. Man Tam, Z. Qi, Y. Jia, Dmitri K. Efetov, Oskar Vafek, Nicolas Regnault, Hongming Weng, Q. Wu, B. A. Bernevig, J. Yu,
 Moiré Fractional Chern Insulators II: First-principles Calculations and Continuum Models of Rhombohedral Graphene Superlattices, Phys. Rev. B. \textbf{109}, 205122 (2024).
 
 \bibitem{Liu2023}
 Z. Guo, X. Lu, B. Xie, J. Liu,
 Theory of fractional Chern insulator states in pentalayer graphene moiré superlattice, arXiv:2311.14368.
 
 \bibitem{Parker2024}
 T. Soejima, J. Dong, T. Wang, T. Wang, M. P. Zaletel, A. Vishwanath, D. E. Parker,
 Anomalous Hall Crystals in Rhombohedral Multilayer Graphene II: General Mechanism and a Minimal Model, arXiv:2403.05522.
 
 \bibitem{BernevigPrivate}
B. A. Bernevig, private communication; X. Li, private communication.
 
\bibitem{Mott1969}
N. F. Mott,
Conduction in non-crystalline materials, Philosophical Magazine
\textbf{19}, 835 (1969).

\bibitem{ES1975}
A. L. Efros and B. I. Shklovskii, 
Coulomb Gap and Low Temperature Conductivity of Disordered Systems. 
J. Phys. C \textbf{8}, L49 (1975).

\bibitem{footnote1}
The extracted gaps depend somewhat, but not very strongly, on the temperature window of the fits and 
we have chosen the broadest fit windows for our results.

\bibitem{West1993}
R. R. Du, H. L. Stormer, D. C. Tsui, L. N. Pfeiffer, and K. W. West,
Experimental evidence for new particles in the fractional quantum Hall effect,
Phys. Rev. Lett. \textbf{70}, 2944 (1993).

\bibitem{Shayegen2022}
K. A. Villegas Rosales, P. T. Madathil, Y. J. Chung, L. N. Pfeiffer, K. W. West, K. W. Baldwin, and M. Shayegan,
Composite fermion mass: Experimental measurements in ultrahigh quality two-dimensional electron systems,
Phys. Rev. B \textbf{106}, L041301 (2022).

\bibitem{Shayegen2021}
K. A. Villegas Rosales, P. T. Madathil, Y. J. Chung, L. N. Pfeiffer, K. W. West, K. W. Baldwin, and M. Shayegan,
Fractional Quantum Hall Effect Energy Gaps: Role of Electron Layer Thickness,
Phys. Rev. Lett. \textbf{127}, 056801 (2021).

\bibitem{Morf2002}
R. H. Morf, N. d’Ambrumenil, and S. Das Sarma,
Excitation gaps in fractional quantum Hall states: An exact diagonalization study,
Phys. Rev. B \textbf{66}, 075408 (2002).

\bibitem{Jain2002}
V. W. Scarola, S.-Y. Lee, and J. K. Jain,
Excitation gaps of incompressible composite fermion states: Approach to the Fermi seam,
Phys. Rev. B \textbf{66}, 155320 (2002).

\bibitem{HalperinBook}
B. I. Halperin in \textit{Perspectives in Quantum Hall Effects: Novel Quantum Liquids in Low-Dimensional Semiconductor Structures}, 
edited by S. Das Sarma and A. Pinczuk, 1st ed. (Wiley-VCH, Weinheim, 1996).

\bibitem{JainBook}
J. K. Jain in \textit{Perspectives in Quantum Hall Effects: Novel Quantum Liquids in Low-Dimensional Semiconductor Structures}, 
edited by S. Das Sarma and A. Pinczuk, 1st ed. (Wiley-VCH, Weinheim, 1996).

\bibitem{Barkeshli2012}
M. Barkeshli and X.-L. Qi,
Topological Nematic States and Non-Abelian Lattice Dislocations,
Phys. Rev. X \textbf{2}, 031013 (2012).

\bibitem{Barkeshli2021}
N. Manjunath and M. Barkeshli,
Crystalline gauge fields and quantized discrete geometric response for Abelian topological phases with lattice symmetry,
Phys. Rev. Research \textbf{3}, 013040 (2021).


\end{thebibliography}
\end{document}